\documentclass[preprint]{aastex}

\shortauthors{Gizis and Harvin}
\shorttitle{L Subdwarfs}

\begin{document}

\title{Halo Stars Near the Hydrogen-Burning Limit: The M/L Subdwarf Transition\footnote{Observations reported here were obtained at the MMT Observatory, a joint facility of the University of Arizona and the Smithsonian Institution.}}

\author{John E. Gizis, James Harvin}
\affil{Department of Physics and Astronomy, University of Delaware, 
Newark, DE 19716}

\begin{abstract}
We present the first far-red spectra of the L subdwarf 2MASS J16262034+3925190 and the late M/early L subdwarf 2MASS J16403197+1231068.  We confirm the ultracool subdwarf nature of these objects.  2M1626+3925 shows strong K~I absorption, like an L4 dwarf,  and is very similar to, but hotter than, the late L subdwarf 2MASS J05325346+8246465.  It is unambiguously an L subdwarf.  2M1640+1231 is very similar to SSSPM J1444-2019, which has been classified as sdM9 or early sdL.   In contrast to the hotter M subdwarfs, L subdwarfs are  characterized by not only enhanced hydrides but also strong TiO.  Progress in a classification system requires identification of more subdwarfs to map out their variations.   
\end{abstract}


\section{Introduction}

Although most stars in the solar neighborhood are members of the Galactic disk, there is a trace population of old, low-metallicity stars.   The spectra of near-solar metallicity low-mass stars can be classified in the M dwarf scheme \citep{khm} and are dominated in the red by molecular absorption, most notably TiO.   The spectra of metal-poor low-mass stars are dramatically different, and can be classified in the scheme of \citet{g97}[hereafter G97], who divided the stars into M dwarfs (M), subdwarfs (sdM) and extreme subdwarfs (esdM), on the basis of the TiO and CaH features.  The sdM have reduced but significant TiO absorption compared to M dwarfs, while extreme M subdwarfs have weak or even absent TiO.  At and below the hydrogen-burning limit, solar metallicity dwarf spectra begin to look qualitatively different, and are described by the L dwarf classification scheme \citep{k99}.  TiO, and later VO, weaken and disappear as dust grains form; the neutral alkali (Na, K, Rb, Cs) lines are strong; and the hydrides strengthen (see Kirkpatrick 2006 for a complete review and references.)  

Although the G97 system covers the LHS stars \citep{l79,faintlhs2}, new surveys have begun to identify even cooler subdwarfs  and have extended the sdM and esdM sequences \citep{esdm7,sdm8,esdm65, sdm95,esdm8} to later spectral types.   Four subdwarfs have merited suggestions that they should be considered L subdwarfs by their discoverers.   \citet{latesdl} found 2MASS J05325346+8246465\footnote{Hereafter 2M0532+8246; we follow this convention for abbreviation for other object designations.} as part of a T dwarf search ---  it is dramatically different in both the red and near-IR from any L dwarf and is most likely a metal-poor brown dwarf.  \citet{earlysdl}  discovered LSR1610-0040 as part of a photographic plate-based proper motion survey --- it resembles an sdM but has anomalously strong Rb~I and Cs~I lines, leading to the suggestions that it should be considered an early L dwarf.  However, both \citet{c1610} and \citet{b1610} have concluded it is peculiar mid-M subdwarf with unusual abundances.  Most recently, \citet{sss1444} report "the formal classification of SSSPM J1444-2019 is sdM9, although it appears likely that it is, in fact, an L-type subdwarf."   \citet{latesdl2} reports 2MASS J16262034+3925190 as an L subdwarf that is not as cool as 2M0532+8246

We present red spectra of three cool subdwarfs:  LSR1610-0040, 2M1626+3925, and 2MASS J16403197+1231068.  Like 2M1626+3925, 2M1640+1231\citep{bother} was discovered in a 2MASS-based T dwarf search.  It  is classified as "sdM8?" on the basis of a near-infrared spectrum, but existing "sdM9" type objects have been classified in the red.  We present the data in Section~2 and discuss issues in the classification and interpretation of these objects in Section~3.  

\section{Data}

We obtained red spectra covering the useful range 5500 to 9000 \AA~ at a resolution of 11\AA~ using the MMT Red Channel Spectrograph on the night of 7 September 2004.  The spectra were extracted and calibrated using IRAF in the usual way.   The spectra and some of the major features used to classify are shown in Figure 1.  There appears to be some fringing  redward of 8000\AA.  In Table 1 we list our measurements of selected G97 and \citet{k99} spectral features. Note that the CaH2, CaH3, and TiO5 indices used by G97 are smaller ($<1.0$)  for more absorption, while the CrH-a and FeH-a indices from \citet{k99} are larger ($>1.0$) for  more absorption.   We estimate the uncertainty on our bandstrengths indices at $\pm 0.02$; the LSR1610-0040 measurements are indeed consistent at this level with \citet{earlysdl}.   We use these indices to guide our discussion of the appearance of the features in the next section; there are not yet enough of these objects to populate index-index plots.  As in the existing sdM sequence \citep{faintlhs2}, there is no H$\alpha$ emission in any of these subdwarfs.

\section{Discussion}

Spectral classification is a powerful scientific method that aids in the interpretation of stars; 
judgments about the relative appearance of spectra creates a framework in which significant variations can be identified and described.  Ideally, spectral classification is based on the appearance of all the features of a spectrum in a well-defined wavelength region; in modern practice for cool dwarfs, quantitative spectral indices guide the classification or can even be used in formulaic typing once a system is established.   We have a well-established framework for interpreting solar-metallicity M and L dwarf spectra, and an appropriate extension for the M subdwarfs, but we have only a few examples that can be considered L subdwarfs.  It is therefore not appropriate at this time to establish numerical types or the M/L transition.  However, it is appropriate to try to interpret the appearance of these new spectra in the context of the existing framework.

Indeed, \citet{latesdl} argued that that the gross features of 2M0532+8246 could be understood in this way; the spectrum at first glance looks like a L dwarf, but most notably has TiO features, enhanced hydrides, and H$_2$ absorption that makes the near-IR colors very blue.   Since TiO disappears in L dwarfs due to the depletion of Ti into grains \citep{jones}, the presence of TiO can be understood as the relative inefficiency of dust formation \citep{b1610}.   (The lack of dust opacity might also contribute to the blue near-IR colors.)  Here we discuss our two targets.

{\bf 2M1640+1231}:  The most striking aspect of the spectrum is its late-M dwarfish appearance but with enhanced hydrides; this is consistent with the "sdM8?" near-IR classification.  The 7100-7300 \AA~ region shows very strong TiO absorption.   This feature actually weakens in M8 and later dwarfs, but here it is somewhat stronger than even in an M7 dwarf.  On the blue side of the 7040\AA~ pseudo-continuum peak, CaH is even stronger than the TiO absorption on the red side; this marks the subdwarf nature of the star.  The TiO and CaH indices confirm this visual impression, marking this as a very late-type star and later than the G97 sdM7.  Na~I is very strong, again pointing to high surface gravity and/or low metallicity.    The 7300-7500 \AA~ region is flat, suggesting the presence of VO absorption; late M and early L dwarfs actually dip in this region \citep{k99}.  The strong dwarf-like features point to an sdM rather than an esdM classification.  In this context, it is worth noting that synthetic model spectra that lacked dust formation and opacity (e.g. Allard \& Hauschildt 1995) showed TiO and CaH absorption continuing to increase at low temperatures with a strong pseudo-continuum.  Overall, 2M1640+1231 seems to be a metal-poor version of an M8-M9 dwarf, with low metallicity resulting in dust grains becoming less important in the spectrum.   

Is a classification of "sdM8" justified by comparison to already published subdwarfs?  2M1640+1231 certainly appears later in spectral type than SSSPM J1013-1356, 
classified as sdM9.5 by \citet{sdm95}, primarily due to the stronger TiO absorption.    It is, in both overall appearance and molecular (TiO5, CaH2, CaH3)  indices, quite similar to SSSPM J1444-2019 \citep{sss1444}.  SSSPM J1444-2019, with the strongest TiO5 absorption yet measured, is evidently slightly cooler, but should belong to the same spectral type.   The optical spectrum can thus best be described as "sdM9" or "early L" following the same arguments in \citet{sss1444}; we discuss this issue further in Section 4.   It is apparent, however, that at least some of the late sdMs can have TiO absorption that are stronger than those observed in disk M dwarfs.

{\bf 2M1626+3925}:  The red spectrum strongly confirms the \citet{latesdl2} L subdwarf near-IR classification, with a somewhat earlier (hotter) type than 2M0532+8246.  There is an overall resemblance to an L4 dwarf (Figure 2), particularly in the appearance of the K~I line cores and broad wings.  (Burgasser et al. noted that 2M0532+8246 resembles an L7).   Nevertheless, numerous features appear different, which we discuss here.  The presence of strong TiO absorption past 7053\AA~ is consistent with \citet{latesdl}'s observations of 2M0532+8246; there is a narrow feature at 8432 \AA~ that is also due to TiO, but it is narrower than in the L dwarfs.   This TiO is also consistent with the strong TiO already noted in the latest sdMs.  The CrH feature strength, as measured by its index,  is consistent with mid-L dwarfs \citep{k99}, but the FeH feature is much stronger than in the mid-L dwarfs.   The atomic lines Rb~I (7800\AA, 7948\AA), Na~I (8183\AA,8195\AA), Cs~I (8521\AA) are present and apparently as strong as in mid-L dwarfs.    We are confident in the detection of these features, but the fringing makes  measurements of their strength impossible.  The CaH features shortward of 7040\AA~are much stronger than any L dwarf.  While the temperature of 2M1626+1231 is unknown, it is clear that this L subdwarf is close to the hydrogen-burning limit for its metallicity since at an age of $>10$ Gyr, it too cool to be higher-mass star and too hot to be a low mass brown dwarf; but even assuming a halo age, an exact identification as a star or a brown dwarf would require reliable measurements of its parallax, bolometric correction, effective temperature, and metallicity and comparison to models.  Lithium is not detected, as expected.  There is a very strong feature near $6570\AA$ which is present in all the individual spectra of this object.  We suggest this is neutral calcium ($6573\AA$), which is also strong in the late sdM plotted in G97.

\section{Summary and Conclusions}

2M1640+1231 can be understood as an subdwarf analog to an M8/M9 dwarf; the current optical classification suggests a type even later than the sdM8 IR classification -- either sdM9 or early sdL.  It is very similar to SSSPM J1444-2019.  2M1626+3925 can be understood as an L subdwarf; it is similar to, but 'earlier' (hotter) than,  2M0532+8246.   LSR1610-0040 does not appear to be in between these two objects; however, the features pointed out by \citet{earlysdl} are confirmed.   Both \citet{b1610} and \citet{c1610} have extensive discussions of this peculiar object and its classification.  In contrast to the situation in the range sdM0-sdM7, the latest M and L subdwarfs show {\it enhanced} TiO absorption relative to late-M and L dwarf spectra.  Physically, this is due to dust being relatively less important than in solar-metallicity dwarfs, as suggested by \citet{latesdl}.

There are two important points that emerge from the difficulty in dealing with the spectra of these stars.
First,  if the "L subdwarf" term is to suggest features in common with the L dwarf sequence, both 2M1640+1231 and LSR1610-0040, as well as SSSPM J1013-1356 and SSSPM J1444-2019, should be considered M subdwarfs.  However, as noted by \citet{earlysdl} and \citet{sss1444}, we have run out of single digit sdM types, especially if SSSPM J1013-1356 is an sdM9.5, pointing to an early L designation.
The G97 system has been successful in distinguishing the dwarf, subdwarf and extreme subdwarfs bins, and in its numerical types it distinguishes within these bins.
Types were assigned so that CaH absorption was similar for all M, sdM and esdM; for the hotter stars recent work confirms that the CaH is primarily sensitive to temperature \citep{ww06}, but since the latest types have CaH stronger than in M dwarfs this system necessarily breaks down.  LHS 377 is {\it by definition} an sdM7 in the G97 system, but it was the only object later than sdM5 known at that time!  At least in the red spectrum here, an sdM9 classification of 2M1640+1231 -- the same as SSSPM J1444-2019 -- suggests that existing numerical types are not wildly inappropriate.  It is therefore possible that the community may prefer to describe such objects as SSSPM J1444-2019 and 2M1640+1231 as sdM9, and therefore compress the existing late sdMs that have extended the sdM sequence into the sdM6-sdM8 subclasses.   
The community could alternatively allow the sdM spectral classes to extend beyond sdM9; e.g., sdM10, sdM11, etc., if it is desirable to avoid sdL types.   
The behavior of TiO and CaH suggests that in subdwarfs that the transition from M to L is rather different at low metallicity, so a one-to-one matching may not be a good description of the spectra anyway: 2M1626+3925 is dominated by the same CaH and TiO features as sdMs in the limited range 6500-7300\AA.  Furthermore, the 2MASS L subdwarfs are {\it blue} in  $J-K_s$, yet L dwarfs are much {\it redder} in $J-K_s$ than M dwarfs.     The change in the overall appearance once enough subdwarfs have been discovered should govern the M/L transition, and is likely to be better linked to physical changes.    It may well be, that happened with T dwarfs, that near-infrared classification is more useful:  see \citet{esdm8} for just such a suggestion.  
The best solution is to wait until more late sdM/early sdL's have been discovered and the relative merits of red and near-IR typing can be assessed.  

Second, the fact that the three objects presented in this paper seem to be an extension of the sdM sequence, rather than the esdM sequence, yet do not clearly fit into a sequence strongly suggests that we need more objects simply to characterize the appearance of the spectra near the M/L transition and that an extra parameter may be needed to describe them.  Such factors as unresolved binaries, changes in $[\alpha/Fe]$ abundances,  and higher sensitivity to $[m/H]$ might all play a role.  The fact that an object,   LSR1610-0040, that is so difficult to fit into the existing framework \citep{earlysdl,c1610,b1610}  has already been found should serve as a cautionary warning.  \citet{sss1449}'s suggestion that the latest sdM and sdL's are only $[m/H]=-0.5$ is also consistent with the idea that the coolest dwarfs are very sensitive to metallicity; on the other hand, it seems to us, following G97, that if $[m/H]=-0.5$ caused subdwarf classification then there would be large numbers of the them already discovered in the disk, rather than only a handful of objects with halo kinematics.  Rather, we believe these objects are likely to have $[m/H] <-0.5$ (i.e., $[m/H] \approx -1$).  Detailed studies of the objects discussed in this paper, as well as such objects as  the blue M7 LHS 1317 \citep{faintlhs2}, are needed to answer these questions.  In any case, we suggest that all numerical spectral subtypes of sdM7 and later, be considered tentative until a large sample with both optical and near-IR spectra can be characterized.  

\acknowledgments

We thank James Liebert for his assistance in observing at the MMT.   We thank the MMT and NOAO staff for their support.  We were allowed use of the MMT through Community Access Time made possible by funding by the National Science Foundation.    Research has benefitted from the M, L, and T dwarf compendium housed at DwarfArchives.org and maintained by Chris Gelino, Davy Kirkpatrick, and Adam Burgasser.

\begin{deluxetable}{lcccccl}
\tablenum{1}
\tablecaption{Spectroscopic Indices}
\tablewidth{0pt}
\tablehead{
\colhead{Target} & \colhead{TiO5} & \colhead{CaH2} & \colhead{CaH3} &
\colhead{CrH-a} & \colhead{FeH-a} & {Source} 
}
\startdata
LSR1610-0040 & 0.28 & 0.28 & 0.49 & 1.09 & 1.19 & This paper\\
2M1626+3925 & 0.25 & 0.10 & 0.13 & 1.68 & 1.91 &  This paper\\
2M1640+1231 & 0.11 & 0.09 & 0.25 & 1.37 & 1.66 & This paper\\
SSSPM J1444-2019 & 0.08 & 0.11 & 0.23 & \nodata & \nodata & \citet{sss1444} \\
SSSPM J1013-1356 & 0.21 & 0.12 & 0.20 & \nodata & \nodata & \citet{sdm95} \\
\enddata
\end{deluxetable}

\begin{figure}
\epsscale{0.7}
\plotone{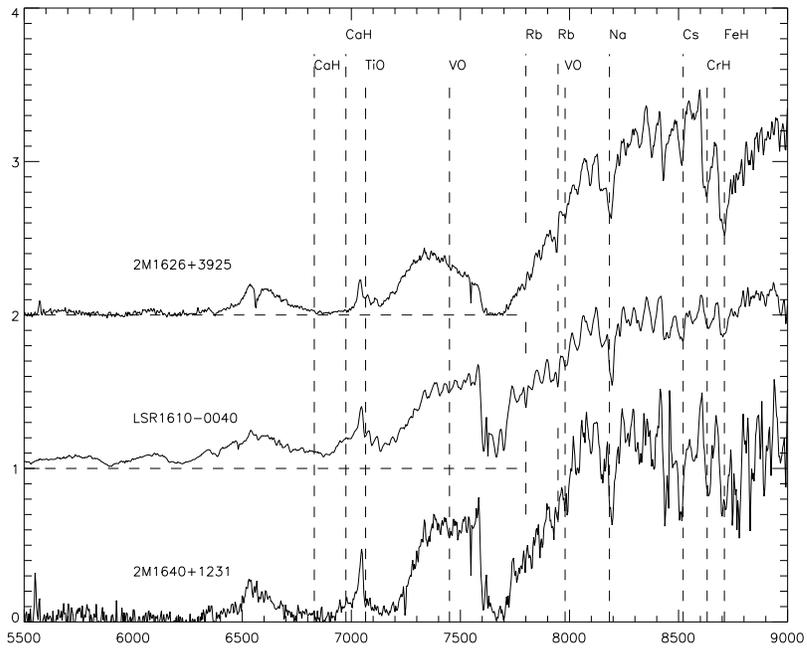}
\caption{The spectra of 2M1640+1231, LSR 1610-0040, and 2M1626+3925.   Each is offset by 1.0 to avoid overlapping.  The dashed line from 5500\AA~ to 7800\AA~ indicates the zeropoint for the two offset spectra.  
\label{fig1}}
\end{figure}

\begin{figure}
\epsscale{0.7}
\plotone{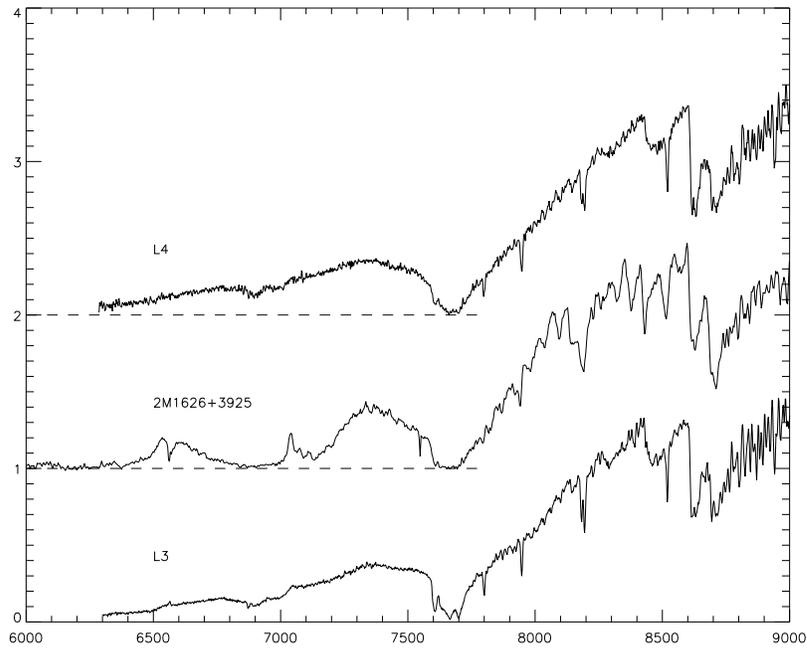}
\caption{The spectra of  2M1626+3925 compared to standard L3 and L4 dwarf Keck spectra from \citet{k99}.  There is a resemblance to an L4 dwarf but the hydrides and TiO are very strong.    The dashed line from 5500\AA~ to 7800\AA~ indicates the zeropoint for the two offset spectra.  
\label{fig2}}
\end{figure}


\begin{thebibliography}

\bibitem[Allard and Hauschildt(1995)]{ah95} Allard, F., \& Hauschildt, P.H.
    1995, \apj, 445, 433

\bibitem[Burgasser et al.(2003)]{latesdl} Burgasser, A.J., et al. 2003, \apj, 592, 1186

\bibitem[Burgasser(2004)]{latesdl2} Burgasser, A.J. 2004, \apj, 614, L73

\bibitem[Burgasser et al.(2004)]{bother} Burgasser, A.J.,McElwain, M.W., Kirkpatrick, J.D., Cruz, K.L., Tinney, C.G., \& Reid, I.N.  2004, \aj, 127, 2856

\bibitem[Burgasser \& Kirkpatrick(2006)]{esdm8} Burgasser, A.J., Kirkpatrick, J.D.  2006, \apj, in press (astro-ph/0603382)

\bibitem[Cushing \& Vacca(2006)]{c1610} Cushing, M.C., \& Vacca, W.D.  2006, \aj, 131, 1797

\bibitem[Gizis(1997)]{g97} Gizis, J.E.  1997, \aj, 113, 806 [G97]

\bibitem[Jones \& Tsuji(1997)]{jones} Jones, H.R.A., \& Tsuji, T.  1997, \apj, 480, L39

\bibitem[Kirkpatrick et al.(1991)]{khm} Kirkpatrick, J.D., Henry, T.J.,
    McCarthy, D.W.  1991, \apjs, 77, 417 

\bibitem[Kirkpatrick et al.(1999)]{k99}  Kirkpatrick, J.D., Reid, I.N., 
Liebert, J., Cutri, R.M., Nelson, B., Beichman, C.A., Dahn, C.C., 
Monet, D.G., Gizis, J.E., \& Skrutskie, M.F.  1999, \apj, 519, 802 

\bibitem[Kirkpatrick(2006)]{k06} Kirkpatrick, J.D.  2006, \araa, 43, 195

\bibitem[L{\'e}pine et al.(2003a)]{sdm8} L{\'e}pine, S., Shara, M.M., \& Rich, R.M.  2003a, \apj, 585, L69

\bibitem[L{\'e}pine et al.(2003b)]{earlysdl} L{\'e}pine, S., Rich, R.M., \& Shara, M.M.  2003b, \apj, 591, L49

\bibitem[L{\'e}pine et al.(2004)]{esdm65} L{\'e}pine, S., Shara, M.M., \& Rich, R.M.  2004, \apj, 602, L125

\bibitem[Luyten(1979)]{l79} Luyten, W.J.  1979, Catalogue of Stars With Proper Motions Exceeding 0.''5 Annually (LHS), (University of Minnesota, Minneapolis, Minnesota)

\bibitem[Reid \& Gizis(2005)]{faintlhs2} Reid, I.N., \& Gizis, J.E.  2005, \pasp, 117, 676

\bibitem[Reiners \& Basri(2006)]{b1610} Reiners, A., \& Basri, G.  2006, \aj, 131, 1806

\bibitem[Schweitzer et al.(1999)]{esdm7} Schweitzer, A., Scholz, R.-D., Stauffer, J., Irwin, M., \& McCaughrean, M.J.  1999, \aap, 350, L62

\bibitem[Scholz et al.(2004a)]{sdm95} Scholz, R.-D., Lehmann, I., Matute, I., \& Zinnecker, H.  2004a, \aap, 425, 519

\bibitem[Scholz et al.(2004b)]{sss1444} Scholz, R.-D., Lodieu, N., \& McCaughrean, M.  2004b, \aap, 428, L25

\bibitem[Woolf \& Wallerstein(2006)]{ww06} Woolf, V.M., \& Wallerstein, G.  2006, \pasp, 118, 218

\end{thebibliography}
\end{document}